\documentclass[mathleft
]{an}
\usepackage{graphicx}
\usepackage{times}
\def\ltsima{$\; \buildrel < \over \sim \;$}
\def\simlt{\lower.5ex\hbox{\ltsima}}
\def\gtsima{$\; \buildrel > \over \sim \;$}
\def\simgt{\lower.5ex\hbox{\gtsima}}
\def\gsimeq
{\hbox{\raise0.5ex\hbox{$>\lower1.06ex\hbox{$\kern-1.07em{\sim}$}$}}}
\def\lsimeq
{\hbox{\raise0.5ex\hbox{$<\lower1.06ex\hbox{$\kern-1.07em{\sim}$}$}}}

\def\fevc{Fe~{\sc xxv}}
\def\fevs{Fe~{\sc xxvi}}

\def\xmm{{\it XMM-Newton }}

\def\exosat{{\it EXOSAT}}

\def\xmm{{\it XMM-Newton}}
\def\chandra{{\it Chandra}}
\def\suzaku{{\it Suzaku}}
\def\rxte{{\it RXTE}}

\def\nustar{{\it NuSTAR}}

\def\apj{ApJ}

\def\mnras{MNRAS}
\def\aap{A\&A}

\def\apjl{ApJ}
\def\apjs{ApJS}
\def\araa{ARA\&A}

\def\nat{Nature}
\def\iaucirc{IAU~circular}
\def\ssr{SSRv}
\def\nar{NewAR}

\def\gros{GROJ1655-40}

\def\axj{AX~J1745.6-2901}
\def\exo{EXO~0748-676}

\def\sgras{Sgr~A$^\star$}

\def\xis{XIS}
\def\xis1{XIS1}
\def\xis2{XIS2}
\def\xis3{XIS3}

\begin{document}

\Pagespan{789}{}% Document's page range. 
\Yearpublication{2006}%
\Yearsubmission{2005}%
\Month{11}%   
\Volume{999}%  
\Issue{88}% 
% \DOI{This.is/not.aDOI}% 

\title{High ionisation absorption in low mass X-ray binaries}

\author{G. Ponti,\inst{1}\fnmsep\thanks{Corresponding author:
  \email{ponti@mpe.mpg.de}\newline}
  S. Bianchi\inst{2}, T. Mu\~{n}oz-Darias\inst{3}, K. De\inst{1,4}, R. Fender\inst{5} \and A. Merloni\inst{1}
}
\titlerunning{Fe~K absorption in LMXB}
\authorrunning{G. Ponti et al.}

\institute{Max Planck Institut f\"{u}r Extraterrestrische Physik, 85748, Garching, Germany
\and
Dipartimento di Matematica e Fisica, Universit\`a Roma Tre, Via della Vasca Navale 84, I-00146, Roma, Italy
\and
Departamento de astrof\'isica, Univ. de La Laguna, E-38206 La Laguna, Tenerife, Spain
\and 
Indian Institute of Science. Bangalore - 560012, India
\and 
Department of Physics, University of Oxford, Denys Wilkinson Building, 
Keble Road, Oxford OX1 3RH UK
}

\received{1 Sep 2015}
\accepted{30 Sep 2015}
\publonline{later}


\abstract{The advent of the new generation of X-ray telescopes 
yielded a significant step forward in our understanding of ionised 
absorption generated in the accretion discs of X-ray binaries. 
It has become evident that these relatively weak and narrow 
absorption features, sporadically present in the X-ray spectra 
of some systems, are actually the signature of equatorial outflows, 
which might carry away more matter than that being accreted. 
Therefore, they play a major role in the accretion phenomenon. 
These outflows (or ionised atmospheres) are ubiquitous 
during the softer states but absent during the power-law 
dominated, hard states, suggesting a strong link with 
the state of the inner accretion disc, presence of the radio-jet 
and the properties of the central source. 
Here, we discuss the current understanding of this field. }

\maketitle

\section{Introduction}

Located at typical distances of few kilo-parsecs and with luminosities 
peaking at $L\sim10^{38-39}$~erg~s$^{-1}$, X-ray binaries are within 
the brightest X-ray sources of the sky, with the more extreme cases 
reaching fluxes of $F\sim10^{-7}$~erg~cm$^{-2}$~s$^{-1}$ (Bird et al. 2004; 2007; 
Remillard \& McClintock 2006; Krivonos et al. 2007). 
Indeed, not surprisingly, the first extrasolar X-ray sources discovered 
were accreting X-ray binaries (Giacconi et al. 1962; 1974). 
These systems are composed of a compact object, either a neutron star (NS) 
or a black hole (BH), orbiting around a normal companion star from which 
they accrete material (van Paradijs 1983). 
The powerful X-ray emission originates from the accretion of such material 
that, because of the angular momentum, forms an accretion 
disc around the compact object (Shakura \& Sunyaev 1973). 

The mass of the companion star divides X-ray binaries in two main classes,  
the high mass and the low mass X-ray binaries (HMXB and LMXB, respectively). 
The geometry of the mass transfer from the companion star is one of the main 
macroscopic differences (of particular relevance here) between the two classes. 
In fact, HMXB accrete most of the material through the wind from the companion star, 
therefore the environment of such systems is filled with flows of highly and lowly 
ionised material from the companion star wind (Lubow \& Shu 1975; Frank et al. 2002). 
The accretion in LMXB, instead, happens through Roche lobe overflow (Frank et 
al. 2002). The companion star fills its Roche lobe. Therefore, the material of the 
companion star at the internal Lagrangian point is in an unstable equilibrium and, 
in part, falls toward the compact object, generating a flow with high angular 
momentum (Lubow \& Shu 1975; Shakura \& Sunyaev 1973; Frank et al. 2002). 
Because of the high angular momentum of this material, an accretion disc 
extending up to a significant fraction of the Roche lobe is therefore formed. 

The wind of the companion stars in HMXB typically generates an 
environment filled with outflowing plasma, potentially complicating the study 
of accretion disc winds. For this reason, most of the works on disc winds, 
that will be discussed here, have been performed in LMXB. 
Indeed, in LMXB the mass transfer from the companion star is supposed to be
confined to the binary orbital plane and to be lowly ionised, leaving the other 
line of sights free from this confusing material. 

\subsection{Dipping phenomenon}

\begin{figure} [ht]
\includegraphics[width=83mm,height=49mm]{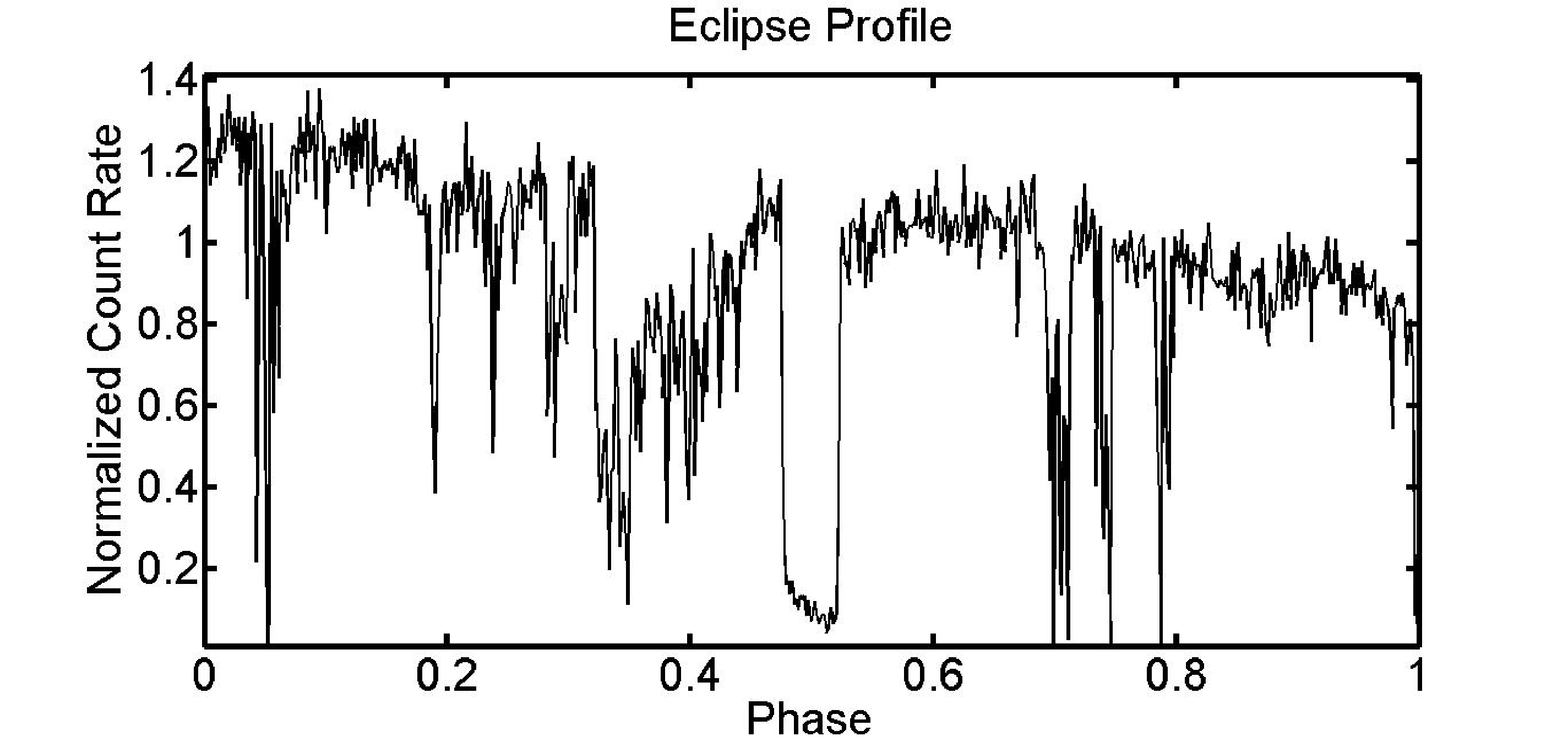}
\includegraphics[width=83mm,height=49mm]{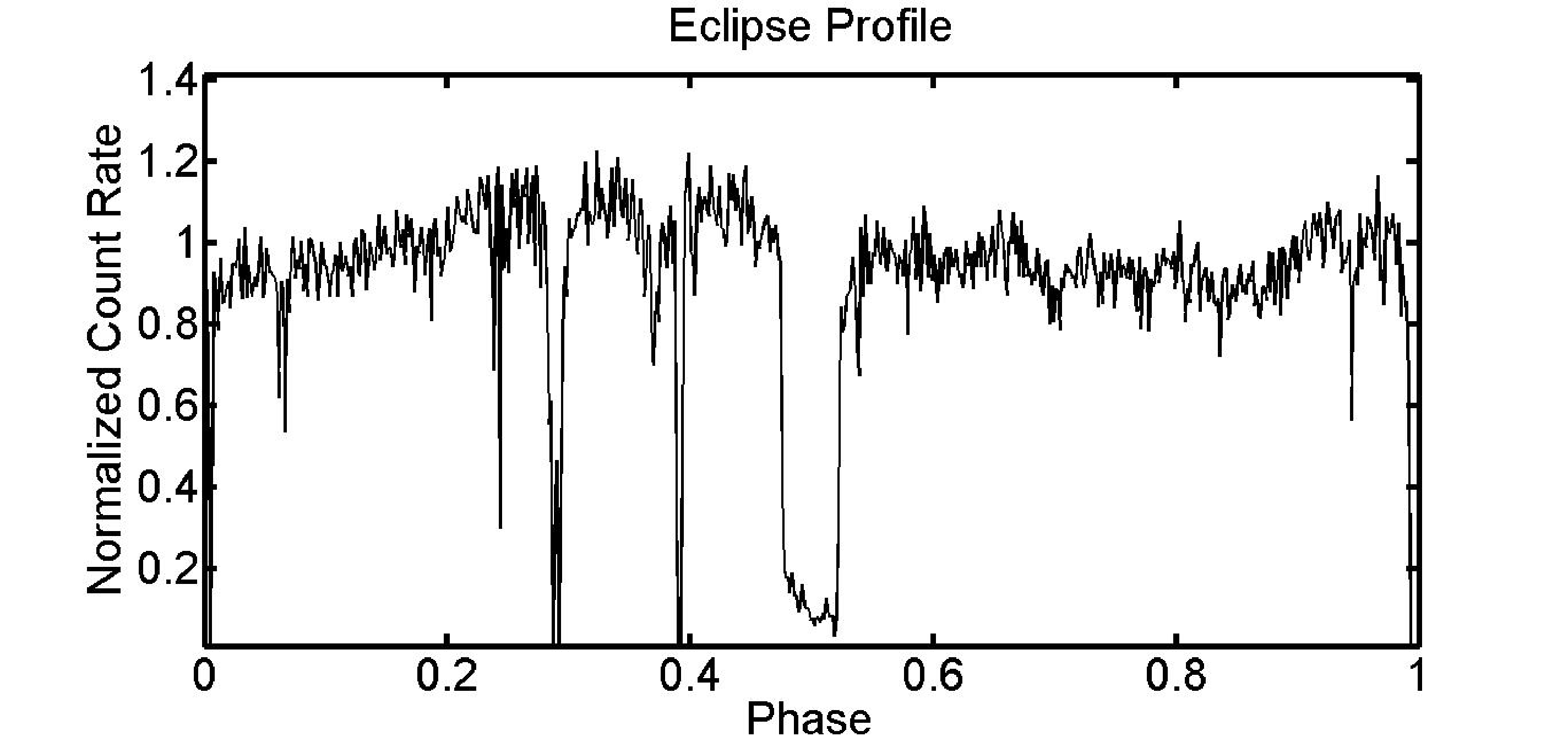}
\includegraphics[width=83mm,height=49mm]{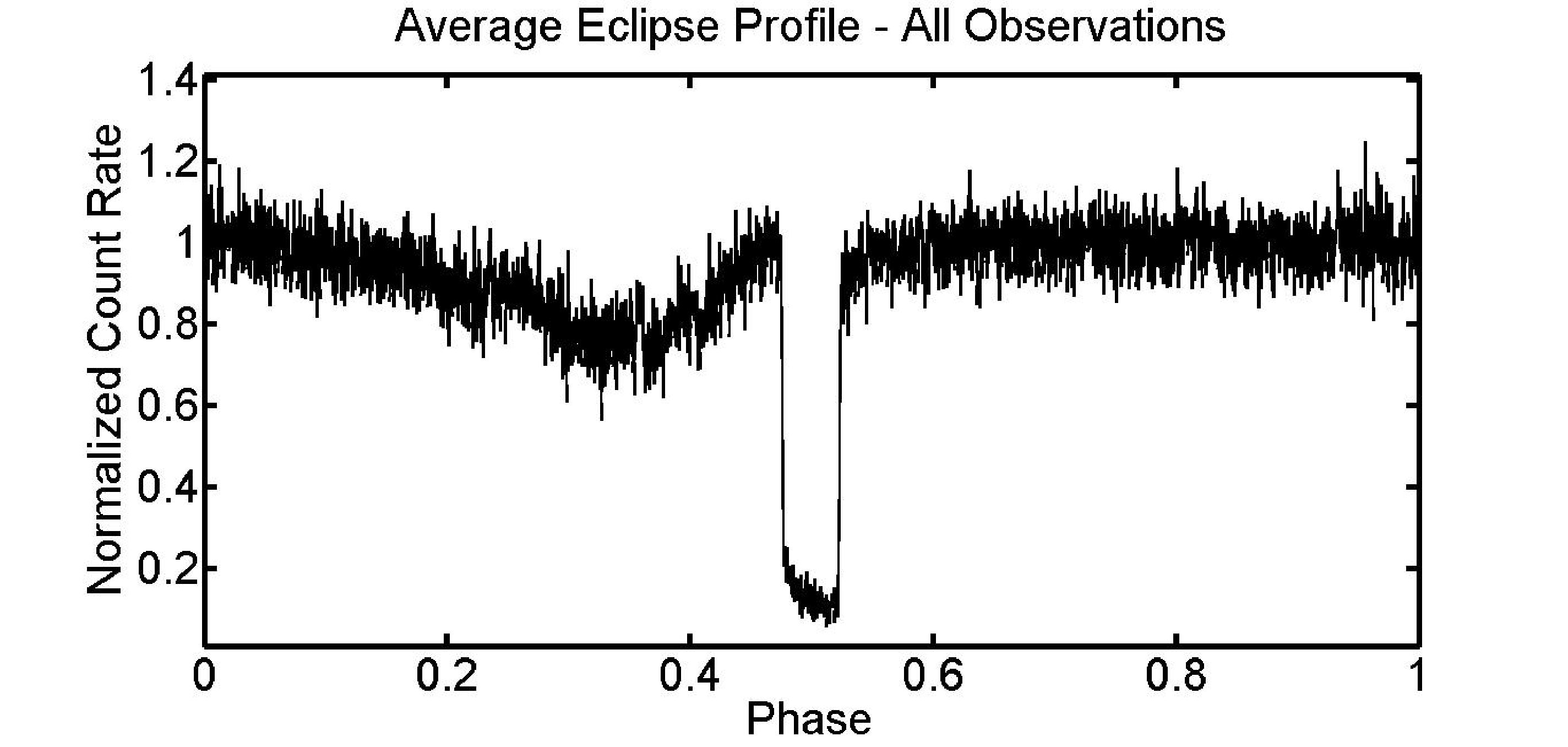}
\caption{{\it (Top and middle panels)} \xmm\ light curve of \axj\ 
during two consecutive orbits, in the 3-10~keV band with 50~s 
time bins. The orbital phase is defined so that the mid eclipse 
happens at phase 0.5. Strong dipping activity is observed during 
one orbit, while almost no dipping is present in the following 
one. {\it (Bottom panel)} Stacked \xmm\ light curve of all the orbits 
of \axj\ available in the \xmm\ archive (for a total of 29). Although 
sporadic, the dipping phenomenon is periodic, peaking between 
orbital phases $\sim0.2$ and $0.45$. }
\label{dip}
\end{figure}
The top and middle panels of Fig. \ref{dip} show the 3-10~keV light curve 
(with 50~s time bins) of \axj\ during two consecutive orbits, as a 
function of the orbital phase. 
\axj\ is an eclipsing (therefore high inclination) neutron star LMXB. 
All light curves in Fig. \ref{dip} show the presence of a drop at the time 
of the eclipse, at orbital phase 0.5. Moreover, the most evident 
features in the top panel of Fig. \ref{dip} are many narrow and 
intese drops of the source count rate, the so called dipping 
phenomenon. Such features almost completely disappear during 
the following orbit, suggesting that indeed they are sporadic and 
highly variable from orbit to orbit. The bottom panel of Fig. \ref{dip} 
shows the stacked light curve of all the orbits of \axj\ available in the 
\xmm\ archive (see Ponti et al. 2015a,b,c). As typical in 
other sources, we show here for the first time that also for this source 
the dipping phenomenon, although sporadic, is periodic with 
maxima between orbital phase $\sim0.2$ and $0.45$. 

It is well known that the dipping phenomenon is associated to high 
inclination systems (White \& Mason 1985; Frank et al. 1987; 
Diaz-Trigo et al. 2006). It is believed that the turbulence, associated with 
the interaction between the stream of material from the companion star 
and the outer rim of the accretion disc, can substantially alter the height 
of the accretion disc rim with azimuthal angle, generating either 
a thick bulge where the stream hits the disc edge (White \& Mason 1985)
or an inner annulus where the remnant stream circularises (Frank et al. 1987). 
If so, erratic, but periodic, dips would be expected, around orbital phase 
0.2-0.4, in high inclination systems, with maximum elevation of the material 
above the orbital plane of a few tens of degrees (Frank et al. 1987). 

In agreement with this general interpretation, dipping sources are 
observed at high inclination (see e.g. Casares \& Jonker 2014 and 
references therein). Therefore, the dipping phenomenon is a good 
tracer of inclination. 
It has also been confirmed that the dipping phenomenon is produced by 
transient obscuration of the primary X-ray source by a thick layer of 
lowly ionised material\footnote{Having $log(\xi)\lsimeq$3; where $\xi$ is the 
ionisation parameter defined as $\xi=L/(4\pi n R^2)$, where $L$ is the 
source luminosity, $n$ is the absorber density and $R$ the source 
to absorber distance.} (Parmar et al. 1986; Diaz-Trigo et al. 2006). 
Important column densities are typically observed with 
$N_H\gsimeq10^{23}$~cm$^{-2}$ (where $N_H$ is the absorber column 
density) significantly modifying the X-ray spectrum below 2-4~keV 
(Parmar et al. 1986; Diaz-Trigo et al. 2006). 
Also confirmed is that, despite the dips being highly erratic in depth 
and duty cycle, they are recurrent showing periodicities equal 
to the system orbital period (e.g., see bottom panel of Fig. \ref{dip}). 

\subsection{Transient, persistent LMXB and accretion states}

LMXB are known to present fairly different behaviours on long time scales. 
Indeed, some systems are persistently bright, others, instead, are transient 
sources occasionally showing sporadic very luminous outbursts typically 
lasting months to years (sometimes decades). This transient-vs-persistent 
dichotomy is thought to be linked to a thermal-viscous disc instability 
generated by the ionisation of hydrogen (Meyer \& 
Meyer-Hofmeister 1981; Lasota 2001). 
In this model, persistent sources are just like the transients, different 
only because the mass transfer from the companion star is so 
high that the temperature of the accretion disc never falls below the 
threshold for the hydrogen to recombine (Dubus et al. 2001; Coriat et al. 2012) .

Already the first spectral studies of X-ray binaries showed clear evidence 
that LMXB transit through clearly different states during an outburst. 
The initial notion of states in neutron stars was based primarily on 
colour-colour plots and to classifications according to the trail that the source 
draws into such plot, during an outburst (e.g., atols, z-sources). 

The identification of the distinct accretion states, their hysteresis pattern, 
the link to the accretion geometry, optical depth and radiative efficiency 
as well as the link to outflows and patterns of variability has been 
developed first for BH systems (Fender et al. 2004; Belloni et al. 2011).
During an outburst, a LMXB goes through an hysteresis pattern following 
a series of accretion states. Softer states are dominated by: i) the multi-temperature 
disc black body emission peaking around 1~keV (well described 
by an alpha-disc; Shakura \& Sunyaev 1973); ii) low level of variability (integrated 
rms lower than $\sim5$~\%; Mu\~{n}oz-Darias et al. 2011) and; iii) quenched jet 
emission (Fender et al. 1999). During harder states, instead: i) the disc emission 
is significantly weaker (the disc is, possibly, truncated) and the X-ray spectrum 
is dominated by a hard Comptonisation component peaking around $\sim100$~keV; 
ii) the source is highly variable and; iii) a compact jet is always observed 
(Fender et al. 2004; Belloni et al. 2005). 
Interestingly the source moves, during an outburst, through the various states 
following a clear hysteresis pattern, rising in the hard state, eventually transiting 
to the soft state, then declining down to a few percent of the Eddington limit and 
transiting back to the hard state (see Fender \& Belloni 2012, for a recent review).  
Very recently it has been realised that, once the complicating contribution 
from the boundary layer is removed, the same states observed in BH are 
present in NS too (Mu\~{n}oz-Darias et al. 2014). 

\section{Highly ionised (Fe~K) absorption in LMXB} 

\begin{figure}
\includegraphics[width=70mm,height=82mm,angle=-90]{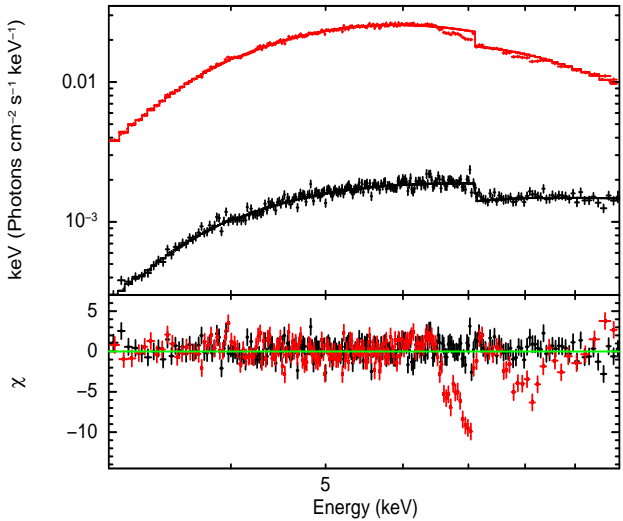}
\caption{\xmm\ X-ray spectra of \axj\ during the soft (red) and hard (black) states. 
Strong features due to ionised absorption (e.g. \fevc\ and \fevs\ 
K$\alpha$ and K$\beta$) are present in the soft state and they 
disappear in the hard state (for more details see Ponti et al. 2015a). }
\label{lines}
\end{figure}
Being located along the Galactic plane, LMXB are typically absorbed 
by significant column densities of Galactic neutral material. 
Indeed, LMXB are generally used as lighthouses to trace 
the distribution of the cold and warm matter in the Galaxy through 
absorption imprinted in their spectra and their dust scattering halo
(Predehl \& Schmitt 1995; Juett et al. 2004; 2006; Costantini et al. 2005). 

In addition to this component, the latest generation of X-ray 
telescopes, equipped with improved spectral resolutions detectors, 
brought to the discovery of narrow absorption lines, due to highly ionised 
material (such as \fevc\ and \fevs), a new component in the spectra 
of LMXB (Brandt \& Schulz 2000; Lee et al. 2002; Parmar et al. 2002; 
Boirin et al. 2004, 2005; Jimenez-Garate, 
Schulz \& Marshall 2003; Ueda et al. 2004; Miller et al. 2006a,b). 
Such absorption lines were observed to be variable, produced 
by very high ionisation plasma and in several cases in outflow. 
All these characteristics indicate that they originate locally in the 
X-ray binary. A very high ionisation state characterise these 
absorbers, with typical values in the range $log(\xi)\sim3.5-5$. 
At such high ionisations, the strongest lines are the K$\alpha$ 
and K$\beta$ transitions of \fevc\ and \fevs\ (see Fig. \ref{lines}
for an example). For this reason, we will refer to this component 
as Fe~K absorption, hereinafter. 

We note that, initially it was not well appreciated the deep link 
between this newly discovered component and the source 
properties and accretion state. Indeed, at a first glance, these 
absorption lines appear as un-impressive narrow (with typical 
broadening of the order of $\sim500-1000$~km~s$^{-1}$ or lower) 
lines with relatively small equivalent widths of $EW\sim10-40$~eV 
(see Fig. \ref{lines}). 
Remarkable similarities are observed between the highly ionised 
absorption observed in NS and BH systems. However, also some 
outstanding differences are seen. 

\subsection{Soft state, equatorial disc winds in BH}

Since the beginning of this century, high resolution spectroscopy 
has allowed to carefully detail the kinematic of highly ionised absorbers. 
Up to the present day, all the high ionisation absorption lines 
in BH systems are observed to be in outflow, therefore they are signatures 
of winds (but see Miller et al. 2014). 
The typical outflow velocities are in the range 
$v_{\rm out}\sim100-2000$~km~s$^{-1}$. 
Interestingly, Fe~K absorption is not observed in all BH systems. 
Thanks to a compilation of all the BH LMXB observed with 
\chandra, \xmm\ and \suzaku, it has been recently realised that 
the Fe~K wind is present only in high inclination systems 
(Ponti et al. 2012). This is a clear evidence that the wind 
has an equatorial geometry and it has a limited covering factor 
(e.g. few tens of degrees above the accretion disc, 
similar to the dipping phenomenon). Indeed, this is in agreement 
with the lack of evidence for the wind re-emission lines, expected
for fully covering absorbers (Lee et al. 2002). 
The link between the dipping phenomenon and the highly ionised 
material appears therefore to be just coincidental (simply the product 
of the high inclination of the system). Indeed, the narrow absorption 
lines are present also during non-dipping periods and, differently 
from the dipping phenomenon, no clear modulation with orbital 
phase is observed (Boirin et al. 2005; Diaz-Trigo et al. 2006; 
Ponti et al. 2014; 2015a). 

One of the most peculiar properties of these winds, that underlines 
the deep link with the inner accretion process, is that they are observed 
primarily and consistently in the so called soft states (Neilsen et al. 2009; 
Ponti et al. 2012). The presence of the wind at all times in the soft state 
indicates a high filling factor, e.g., higher than the one producing the 
dipping phenomenon. For example, if the wind were patchy in the azimuthal 
angle, then we would expect to see some soft state spectra with 
no wind. 
On the other hand, the Fe~K lines, signatures of the wind, disappear 
during the hard state (Ponti et al. 2012). 
The strong connection between winds and source states requires an 
explanation. One obvious reason would be that the wind is over-ionised 
in the hard state. Indeed, when present, the wind appears to increase 
its ionisation state with luminosity, as expected (Ueda et al. 2010; 
Diaz-Trigo et al. 2012; Ponti et al. 2012). However, it has been shown
that over-ionisation fails at explaining the disappearance of the Fe~K 
lines in the hard state, at least in the few cases when detailed 
photo-ionisation computations have been performed (Miller et al. 2012; 
Neilsen et al. 2012; Ponti et al. 2015a). This suggests that disc winds 
are actually missing in the hard state, with the wind being present 
when a standard accretion disc is present and the jet is absent. 

\chandra\ HETG observations of Fe~K winds allow an accurate measurement 
of the wind outflow velocity ($v_{\rm out}$) and a characterisation of the 
ionisation state of the absorber. From these quantities, assuming, from statistical 
arguments, that the wind opening angle is $\sim30^\circ$ (Ponti et al. 2012), 
it is possible to estimate the wind mass outflow rate using the equation 
\begin{equation}
\dot{M}_{wind} = 4\pi m_p v_{\rm out} \frac{L}{\xi} \frac{\Omega}{4\pi}
\end{equation}
where $m_p$ is the proton mass and $\Omega$ is the solid angle 
subtended by the wind. 
The measured values are generally either of the order of, or higher, 
than the mass accretion rates (Lee et al. 2002; Ueda et al. 2004; Neilsen, 
Remillard \& Lee 2011; Ponti et al. 2012). This strongly indicates that these 
winds are a fundamental component in the balance between accretion and 
ejection and they are major ingredients in the accretion process. 
Disregarding such winds would mean overlooking the 
majority of the mass involved in the accretion process. 
In particular, a higher mass transfer rate from the companion star, 
compared to what is generally assumed, might be required. 
This might lead to a more rapid evolution of the binary orbit than we expect. 
Indeed, eclipse timing studies of eclipsing LMXB typically fail to reproduce 
the observed orbital evolution through conservative mass transfer 
(see Ponti et al. 2015d and references therein). 

\begin{figure}
\vskip-0.8cm
\includegraphics[width=73.mm,height=83mm,angle=-90]{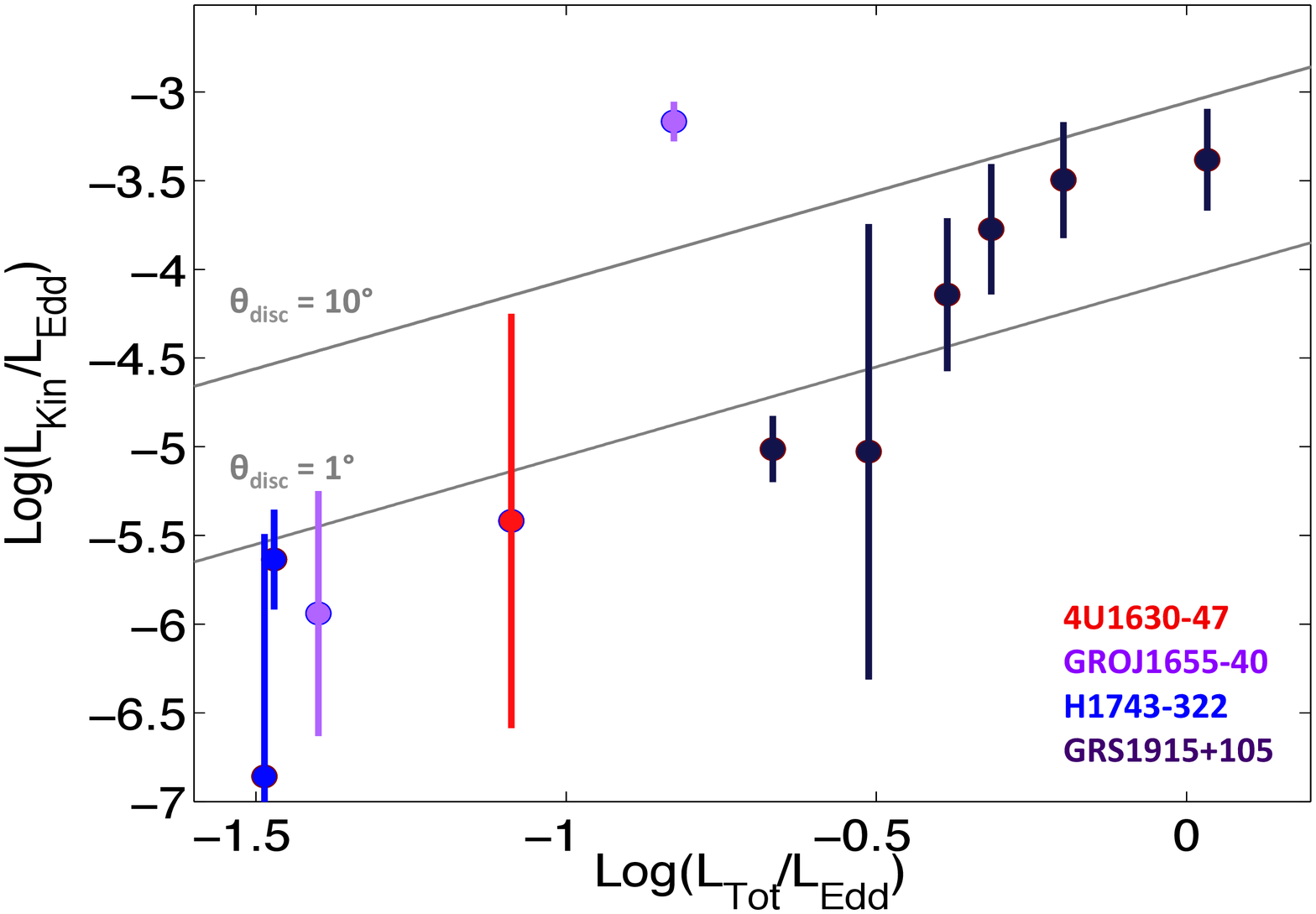}
\caption{Kinetic luminosity of the wind as a function of the source central 
luminosity for several BH LMXB with \chandra\ HETG observations (see 
Ponti et al. 2012 for more details). The grey lines show the expected relation 
for disc flared by $\theta_{\rm disc}=1^{\circ}$ and $10^{\circ}$, assuming a 
conversion between the luminosity intercepted by the disc into kinetic luminosity 
of 0.01. }
\label{Lkin}
\end{figure}

From the mass outflow rate we can compute the wind kinetic luminosity 
as: $L_{kin}=\frac{1}{2}\dot{M}_{wind} v^2_{\rm out}$. 
We note that despite such winds can carry away the majority of the 
transferred mass, their relatively small outflow velocity generates 
very small kinetic luminosities. Indeed, the wind kinetic luminosities 
are observed to span the range $L_k\sim10^{-3.5-6}$ of the Eddington 
luminosity (see Fig. \ref{Lkin}). Even if the same type of rather low velocity 
wind is present in active galactic nuclei, it would have no significant 
contribution for the feedback phenomenon or impact for the evolution of 
the host galaxy (Fabian et al. 2012). Much higher outflow velocities, 
such as the ones observed in ultra fast outflows (Tombesi et al. 2010; 2013)
would be required in this case. 

\subsection{Highly ionised absorption in NS}

NS LMXB also show highly ionised absorption, with a similar range of 
ionisation states and column densities as in BH. Again, as in BH, 
in NS also, ionised absorption is observed only in dipping LMXB, 
suggesting an equatorial geometry (Diaz-Trigo et al. 2006). 

However, there are also outstanding differences. Indeed, it was 
soon realised that two families of highly ionised absorbers are 
present in NS LMXB. In fact, some sources show absorption features 
with outflow velocities of the order of $v_{\rm out}\sim10^{2-3}$~km~s$^{-1}$ 
and are, therefore, called winds, others, show no outflow velocities. 
The ionised structures in these latter sources are sometimes called 
''disc atmospheres''. 

\subsection{A connection between state and Fe~K absorption also in NS?} 

After the discovery of the clear connection between the presence 
of the wind and the accretion state in BH systems (Ponti et al. 2012), 
we investigated if the same connection is present in NS (Ponti et al. 2014; 2015a). 
We first checked the entire sample of NS LMXB with \chandra\ or \xmm\ 
observations (allowing a good characterisation of the properties 
of the Fe~K lines) and we realised that only two sources were standing 
out for their extensive monitoring campaign. One, \exo, has been 
observed more than 20 times because it was a calibration source 
for \xmm, while the second, \axj, has more than 40 observations 
because it falls in the same field of view of \sgras\ (Genzel et al. 2010; 
Ponti et al. 2015b,c). 

\exo\ was discovered by \exosat\ in 1985 (Parmar et al. 1985) and it 
has been active for 23 years, when it finally returned to quiescence 
in 2008 (Hynes \& Jones 2008). \exo\ spent most of the outburst 
in the hard state. Using both the hardness intensity diagrams 
and measuring the source variability, from nearly simultaneous 
\rxte\ observations, Ponti et al. (2014) classified the state of the 
source within each \xmm, \chandra\ and \suzaku\ observation. 
Not one of the 20 X-ray spectra obtained in the hard state reveal 
any significant Fe~K absorption line. On the other hand, intense \fevc\ 
and \fevs\ (as well as rarely observed Fe~{\sc xxiii} plus S, Ar, 
and Ca transitions) lines are clearly detected during the only soft state 
observation. 

Ponti et al. (2015a) analysed all the \xmm\ observations of \axj\ available 
as of 2014 May 14. Eleven observations caught the source in outburst 
of which nine in the soft state and two in the hard state. Significant \fevc\ 
and \fevs\ K$\alpha$ and K$\beta$ lines are observed during all the nine 
soft state observations, while stringent upper limits are observed during the
hard state observations (see Fig. \ref{lines}). The column density 
($N_H\sim2\times10^{23}$~cm$^{-2}$) and ionisation state 
($Log(\xi)\sim4.1$) of the highly ionised absorber are consistent with being 
constant within the soft state observations. Nearly simultaneous \nustar\ 
observations allowed Ponti et al. (2015a) to well characterise the source 
spectral energy distribution both in the soft and hard state. This allowed 
the authors to check that the Fe~K absorption does not disappear because 
of over-ionisation in the hard state. These findings strongly support the 
idea that the same connection between Fe~K absorption and states is 
also valid in the two best monitored NS systems, therefore it is not a 
unique property of BH, but a more general characteristic of accreting 
sources. 

\section{Wind launching mechanisms} 

Many different mechanisms have been invoked to launch winds 
from accreting sources. Among these, some of the most popular 
have at their core: i) the extreme radiation pressure present when 
the source is either at, or over, the Eddington limit; ii) the large 
opacities of the UV transitions generating line driven winds (Castor et al. 1975; 
Murray et al. 1995; Arav et al. 1995) or; iii) the photo-evaporation 
instability producing winds from irradiated molecular clouds (such 
as the torus in AGN; Krolik \& Kriss 2001; Blustin et al. 2005). 
None of these mechanisms is thought to be at work in the case of LMXB. 
Indeed, i) no major difference is typically observed in the wind properties 
of sources spanning a wide range of Eddington ratios, down to 
luminosities of a few per cent Eddington, when the first 
mechanism is expected to be negligible (but see Miller et al. 2006); 
ii) line driven winds are expected to be inhibited by the very 
hard X-ray radiation typical of LMXB, that is very quickly over-ionising 
the plasma (Proga et al. 2002); iii) the reservoir of accreting matter 
is in the companion star and not stored in large molecular clouds, 
therefore also the last mechanism is inhibited. 
LMXB are, therefore, simplified laboratories to study the generation  
of disc winds. 

Only two mechanisms, either magnetic or thermal, are instead 
believed to be able to launch winds in LMXB. 
Thermal (or Compton heated) winds are generated in the outer part 
of a flared accretion disc. At large radii, because of irradiation 
from the central source, the surface of the disc can be heated 
to a point where the thermal velocity exceed the escape velocity. 
In this conditions, an outflow is continuosly generated (Begelman 
et al. 1983a,b; Woods et al. 1996; Luketic et al. 2010). 
The radius at which the thermal velocity is equal to the escape 
velocity is typically referred to as Compton radius.  
An ad hoc configuration of the magnetic field can also, of course, 
generate winds (Blanford \& Payne 1982; see Fukumura et al. 2010; 
2015, for the simulations of magnetic winds in AGN). 
No restriction on either the launching radius or the wind 
outflow velocity is present in this case. 
Therefore any wind in LMXB generated well inside 
the Compton radius is thought to have a magnetic origin. 

\subsection{Magnetic or thermal wind?} 

A significant effort has been undertaken to try to understand if winds 
in LMXB are magnetically or thermally driven. One observation of \gros, 
during the so called anomalous state, shows exceptionally 
intense wind features, with an array of more than 90 absorption 
lines detected at more than 5~$\sigma$ (Miller et al. 2006; 2008; 
Kallman et al. 2009). 
In particular, the detection of the metastable 2s2p$^3$P level of Fe~{\sc xxii} 
implies a narrow range of relatively high number densities 
($n\sim5\times10^{15}$~cm$^{-3}$) for the absorbing plasma. 
Given the observed ionisation state of the absorber and the source 
luminosity, then it can be derived the distance $R$ of the absorber 
by $R=\sqrt{\frac{L}{4\pi n \xi}}\sim5\times10^8$~cm$\sim400$~$r_g$, 
where $r_g=GM_{\rm BH}/c^2$ with $M_{\rm BH}$ being the BH mass, 
$c$ the speed of light and G the gravitational constant. This location 
is well inside the Compton radius. This is a strong argument in favour 
of a magnetic launching mechanism for this wind. We note that in this 
case the wildly employed rule of thumb for which the wind outflow 
velocity is reminiscent of the launching radius of the disc, does not 
hold. Indeed, the observed wind outflow velocity 
$v_{\rm out}\sim300-1500$~km~s$^{-1}$ is characteristic of outer 
regions of the disc (consistent with a thermal origin, although this 
does not exclude a magnetic origin). 
If indeed winds in LMXB have a magnetic origin, then the wind-jet 
anti-correlation (therefore the wind state connection) might be 
easily understood as the product of a reconfiguration of the magnetic 
field lines (first suggested by Neilsen et al. 2009). 
In fact, both jets and winds might be the product of the same 
magnetic outflow. When the magnetic field lines generate a collimated 
outflow, this appears as a jets, while it appears as a wind, 
when un-collimated.  

Apart from the peculiar wind observed during the anomalous state of 
\gros, no other metastable-density-sensitive lines are observed in other 
observations, therefore preventing us from a clear determination of the distance 
to the absorber and therefore to pin down the launching mechanism. 
Apart from a few outliers, typically derived from lines with lower significance, 
the wind outflow velocities are in the range $v_{\rm out}=0$, for disc atmospheres, 
up to $v_{\rm out}=100-2000$~km~s$^{-1}$. These velocities are higher than 
the Keplerian, or local escape velocity, outside $2\times10^4$~$r_g$. 
Therefore these relatively small 
outflow velocities are consistent and suggest a thermal origin for these winds. 
Since their first theorisation, it was predicted that the mass outflow rate in 
thermal winds could be up to many times higher than the mass accretion rate
(Begelman et al 1983a,b). This, also, is in remarkable agreement with observations. 
In thermal winds, the power required to energise the wind is provided by 
the central source. Therefore, the maximum kinetic luminosity of the wind 
is expected not to exceed the central source luminosity intercepted by 
the outer disc and used to heat and launch the wind. 
For discs flared by less than $\sim8-12^{\circ}$, this corresponds to 
a kinetic power of $\sim10^{-3}L/L_{\rm Edd}$ (assuming a conversion 
efficiency of 0.01 from incident luminosity into kinetic power) and expected 
to scale with the source central luminosity. Figure \ref{Lkin} shows that this 
is well consistent with observations. 

\acknowledgements 
GP ackowledge support by the Bundesministerium f\"{u}r Wirtschaft und 
Technologie/Deutsches Zentrum f\"{u}r Luft- und Raumfahrt 
(BMWI/DLR, FKZ 50 OR 1408) and the Max Planck Society.

\end{document}